\documentstyle[aps,pra,floats,epsf]{revtex}
\begin{document}

\twocolumn[\hsize\textwidth\columnwidth\hsize\csname
@twocolumnfalse\endcsname

\title{Restoration of Entanglement by Spectral Filters}
\author{K. Usami$^{1,2}$, Y. Nambu$^{2,3}$, S. Ishizaka$^{2,3}$, T. Hiroshima$^{2,3}$, Y. Tsuda$^{4,5}$, K. Matsumoto$^4$, and K. Nakamura$^{1,2,3}$}
\address{$^1$Depertment of Material Science and Engineering, Tokyo Institute of Technology, 4259 Nagatsuta-chou, Midori-ku, Yokohama, Kanagawa, 226-0026, Japan}\address{$^2$CREST, Japan Science and Technology Corporation (JST), 3-13-11 Shibuya, Shibuya-ku, Tokyo, 150-0002, Japan}
\address{$^3$NEC Fundamental Research Laboratories, 34 Miyukigaoka, Tsukuba, Ibaraki, 305-8501, Japan}
\address{$^4$ERATO, Japan Science and Technology Corporation (JST), 5-28-3, Hongo, Bunkyo-ku, Tokyo, 113-0033, Japan}
\address{$^5$Institute of Mathematics, University of Tsukuba, Tsukuba, Ibaraki, 305-8571, Japan}

\date{\today}

\maketitle

\begin{abstract}
We experimentally demonstrate that entanglement of bi-photon polarization state can be restored by spectral filters. This restoration procedure can be viewed as a new class of quantum eraser, which retrieves entanglement rather than just interference by manipulating an ancillary degree of freedom.
\end{abstract}


\vskip1pc]

\narrowtext

The quantum state which has entanglement is defined that it is composed of
more than two quantum systems and cannot be represented as a mixture of
direct product state of its subsystem\cite{QCQI}. \ Its non-local properties
have been studied by testing the so-called Bell-CHSH inequality\cite%
{CHSH,BellEX}. Entanglement plays the most important role in quantum
information technologies\cite{QCQI}, such as quantum cryptography\cite%
{Crypto}, quantum teleportation\cite{QTtheo,QTInsEX,QTEX}, quantum dense
coding\cite{QDenCode} and quantum computation\cite{QCQI}, which have
recently attracted a great deal of attention. It is known that the more the
states are entangled, the better the state works as a resource for quantum
information processing. It is therefore important to answer the following
questions: how to create the entangled states so as to maximize their
entanglement, how to distill poorly entangled states into highly entangled
states, and how to restore the entanglement of a state that potentially
possesses more entanglement. The third question is the focus of this letter.

Hereafter, we will restrict our consideration to the polarization states of
photons. Several methods to obtain polarization-entangled bi-photon have
been reported\cite{RevESPDC}. Recently, Kwiat et al. devised an easy and
efficient method for producing an arbitrary polarization-entangled state\cite%
{KwiatPRA,WhitePRL,KwiatScience,KwiatNature}. They used cw-pumped
spontaneous parametric down-conversion (SPDC), by which one parent photon in
the cw pump beam is split into two polarization-entangled daughter photons
via {\em two} nonlinear optical crystals, conserving energy and momentum and
satisfying the type-1 phase matching condition\cite{KwiatPRA}. Under
cw-pumped SPDC, the converted photon pair has high purity (low entropy) and
high degree of entanglement, and it can be considered the following Bell
state\cite{KwiatPRA},%
\begin{equation}
\left| \Phi ^{+}\right\rangle =\frac{1}{\sqrt{2}}\left( \left|
H\right\rangle _{A}\left| H\right\rangle _{B}+\left| V\right\rangle
_{A}\left| V\right\rangle _{B}\right) ,  \label{Bell state}
\end{equation}%
where $\left| H\right\rangle (\left| V\right\rangle )$ stand for a
horizontal (vertical) polarized photon state and subscripts A and B
represent two separate observers, Alice and Bob. On the other hand, it has
been reported that under femtosecond pulse-pumped SPDC, the converted photon
pair has poor purity and entanglement, due to the complicated dispersion and
phase matching associated with an ultra-short pump pulse\cite{RevESPDC}.
However, femtosecond pulse-pumped SPDC is indispensable for manipulating
entangled photon {\em pairs}, because of its capability to generate two or
possibly more bi-photons simultaneously\cite{QTInsEX,Zukowski,EnManipuEX}.

In this letter, we report on an experiment of a {\it disentanglement eraser}%
\cite{Garisto-Hardy} which restores entanglement in a bi-photon polarization
state generated by two-crystal, femtosecond-pulse-pumped SPDC. In this
experiment, spectral filters play the role of an effective disentanglement
eraser, which stretch the coherent time of two bi-photon wave packets and
thus erase the information concerning which crystal generates a bi-photon.
By experimentally reconstructing the density matrices, we evaluate how much
entanglement (concurrence\cite{Wootters}) can be restored by a
disentanglement erasing. In order to obtain the reliable density matrices,
we employ two techniques; quantum tomography\cite{WhitePRL,Usami} and a
maximum likelihood method\cite{James}.

\begin{figure}
\epsfxsize=7.0cm
\centerline{\epsfbox{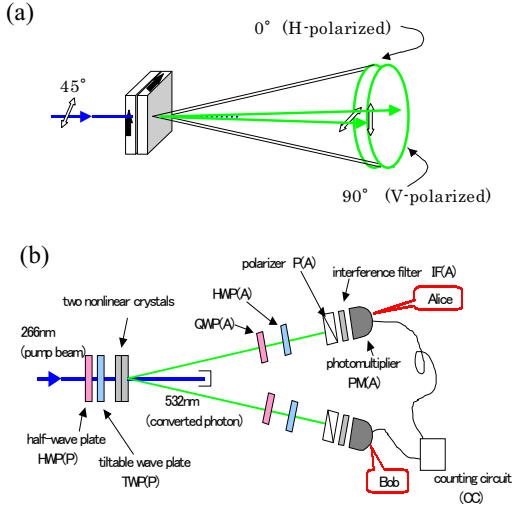}}
\caption{(a) The geometry of two crystals (BBO) for generating
polarization-entangled bi-photons\protect\cite{KwiatPRA}. (b) Experimantal
setup to generate polarization-entangled photon pairs and to evaluate their
entnaglement\protect\cite{WhitePRL,Usami}.}
\label{fig1}
\end{figure}

\begin{figure}
\epsfxsize=7.0cm
\centerline{\epsfbox{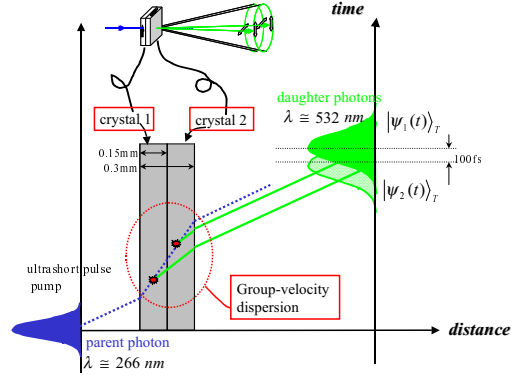}}
\caption{Space-time diagram of two wave packets generated by
femtosecond-pulse-pumped SPDC. In our experiment, the time difference
between two wave packets are about 100fs. This difference is mainly due to
the group-velocity dispertion.}
\label{fig2}
\end{figure}

Our source of entangled bi-photon is almost the same as the source in Ref.%
\cite{KwiatPRA}. Two thin nonlinear crystals, each of which is a 0.15-$%
%TCIMACRO{\unit{mm}}%
%BeginExpansion
\mathop{\rm mm}%
%EndExpansion
$-thick BBO crystal cut for the type-1 phase matching, are attached so that
their optical axes lie in planes perpendicular to each other (Fig.\ref{fig1}%
(a)). The plane including the optical axes of the first (second) crystal and
the propagating direction of the pump beam defines vertical (horizontal). If
the polarization of the pump beam is set to 45$%
%TCIMACRO{\unit{\U{b0}}}%
%BeginExpansion
\mathop{\rm %
{{}^\circ}}%
%EndExpansion
$ to the horizontal, the pump beam state is given by the direct product of
the horizontally polarized and the vertically polarized coherent states with
the same amplitude. Therefore, the following two probability amplitudes can
be ``coherently'' superposed: (1) two horizontally polarized photons
generated in the first crystal or (2) two vertically polarized photons
generated in the second one. Thus, the converted photons can be considered
as a polarization-entangled state as given by Eq.\ref{Bell state}\cite%
{KwiatPRA}. Under femtosecond pulse-pumped SPDC, however, two bi-photon wave
packets, i.e., two probability amplitudes, can be distinguished because of
the considerably short coherent time of the pump pulse\cite{KimPRArapid}. We
will refer to this distinguishability as {\em which-crystal} information.
This which-crystal information usually degrade entanglement of the converted
photon pair. Figure\ref{fig2} shows a space-time diagram, which depicts the
space-time difference between two wave packets. Which-crystal information,
in other words, {\em bi-photon} wave packets appears in the density matrix
as an ancillary degree of freedom, though we are only concerned with the
polarization degree of freedom. We define two bi-photon (temporal) wave
packets, one of which is generated by the first crystal and the another by
the second one, as $\left| \psi _{1}(t)\right\rangle _{T}$ and $\left| \psi
_{2}(t)\right\rangle _{T}$, respectively. They can be expressed as the
inverse Fourier transform, as follows:%
\begin{equation}
\begin{array}{c}
\left| \psi _{1}(t)\right\rangle _{T}=\frac{1}{2\pi }\int d\omega
_{1}g(\omega _{1})e^{-i\omega _{1}t}\left| \omega _{1}\right\rangle \\ 
\left| \psi _{2}(t)\right\rangle _{T}=\left| \psi _{1}(t+\tau )\right\rangle
_{T}=\frac{1}{2\pi }\int d\omega _{2}g(\omega _{2})e^{-i\omega _{2}(t+\tau
)}\left| \omega _{2}\right\rangle .%
\end{array}
\label{Inversetransform}
\end{equation}%
Here we assume that $\left| \psi _{2}(t)\right\rangle _{T}=\left| \psi
_{1}(t+\tau )\right\rangle _{T}$, which means two bi-photon wave packets are
the same, apart from the time difference $\tau $. This time difference is
mainly due to group velocity dispersion (Fig.\ref{fig2}). In Eq.\ref%
{Inversetransform}, $g(\omega _{i})$ and $\left| \omega _{i}\right\rangle $
represent the spectral amplitude and eigenstate of the {\em bi-photon} state
with energy $\hbar \omega _{i}$ for each photon, respectively, and $\left|
\omega _{i}\right\rangle $ constitute a complete orthonormal system (i.e., $%
\int d\omega _{i}\left| \omega _{i}\right\rangle \left\langle \omega
_{i}\right| =1,$ $\left\langle \omega _{j}\right| \omega _{i}\rangle =\delta
(\omega _{i}-\omega _{j})$. Then the state of whole system obtained by
pulse-pumped SPDC can be expressed as%
\begin{equation}
\left| \Omega _{ABT}(t)\right\rangle =\frac{1}{\sqrt{2}}(\left|
H\right\rangle _{A}\left| H\right\rangle _{B}\left| \psi
_{1}(t)\right\rangle _{T}+\left| V\right\rangle _{A}\left| V\right\rangle
_{B}\left| \psi _{2}(t)\right\rangle _{T}).  \label{3 system}
\end{equation}%
What is intriguing about the state of Eq.\ref{3 system} is that its
entanglement can be restored by manipulating the ancillary degree of
freedom, i.e., the bi-photon wave packets, like quantum erasing. Traditional
quantum erasing, e.g., in Young's double-slit configuration\cite{Scully},
involves only two subsystems, principal system A and a ancilla ({\it tag})
system T. In contrast to such a traditional quantum erasing, the state given
by Eq.\ref{3 system} involves three subsystems: two principal systems A and
B, and a tag system T. This framework corresponds to that of a new type of
quantum eraser, i.e., a {\it disentanglement eraser} proposed by Garisto and
Hardy\cite{Garisto-Hardy}, which can be restored {\em entanglement} between
the principal systems A and B, rather than just the interference by a
suitable manipulation of the tag system T. The question now arises: how can
the tag system be manipulated in order to erase which-crystal information?
For erasing which-crystal information marked via a wave packet, we employed
narrow bandwidth spectral filters in front of two photon counting detectors
at the expense of the available flux of entangled photon pairs. Although
there are some reports that the quantum interference visibility in
femtosecond-pulse-pumped SPDC can be restored by erasing the
distinguishability with spectral filters\cite{SpectFilter}, they treat two
principal systems, A and B, as a single subsystem as a whole. Thus, their
eraser can be viewed as a conventional one. Recently, Teklemariam {\it et al.%
} reported the three-spin disentangled eraser on a liquid-state NMR\cite%
{NMR-eraser}. But their effective pure state is separable at any moments of
the processing as mentioned by Braunstein {\it et al.}\cite{Braunstein}.

For evaluating entanglement of a bi-photon generated by pulse-pumped SPDC
with two nonlinear optical crystals, we employed linear tomographic
measurement\cite{WhitePRL,Usami} and a maximum likelihood method\cite{James}%
. Figure\ref{fig1}(b) shows our experimental setup to generate
polarization-entangled photon pairs and to evaluate its entanglement. A
parent photon in the frequency-tripled laser pulse of a mode-locked Ti:
Sapphire laser (central wavelength: 266 $%
%TCIMACRO{\unit{nm}}%
%BeginExpansion
\mathop{\rm nm}%
%EndExpansion
$, pulse duration: \symbol{126}100 $%
%TCIMACRO{\unit{fs}}%
%BeginExpansion
\mathop{\rm fs}%
%EndExpansion
$ and average power: \symbol{126}150 $%
%TCIMACRO{\unit{mW}}%
%BeginExpansion
\mathop{\rm mW}%
%EndExpansion
$) splits into a polarization-entangled photon pair (central wavelength: 532 
$%
%TCIMACRO{\unit{nm}}%
%BeginExpansion
\mathop{\rm nm}%
%EndExpansion
$) via two BBO crystals, as mentioned before. The entangled photons are
emitted into a cone with a half-opening angle of 3.0$%
%TCIMACRO{\unit{\U{b0}}}%
%BeginExpansion
\mathop{\rm %
{{}^\circ}}%
%EndExpansion
$ and detected by photomultipliers (Hamamatsu, H7421-40) with efficiencies
of \symbol{126}40\%. Quarter wave plates (QWP(A) and QWP(B)), half wave
plates (HWP(A) and QWP(B)) and polarizing beam splitters in front of the
detectors, can be used to project the polarization state of down-converted
photons onto any kind of product states of polarization by coincidence
counting measurements (in our experiments, a coincidence window of 6 $%
%TCIMACRO{\unit{ns}}%
%BeginExpansion
\mathop{\rm ns}%
%EndExpansion
$). We chose four projection states for Alice, $\{\left| H\right\rangle
,\left| V\right\rangle ,\left| D\right\rangle =\frac{1}{\sqrt{2}}(\left|
H\right\rangle +\left| V\right\rangle ),\left| R\right\rangle =\frac{1}{%
\sqrt{2}}(\left| H\right\rangle +i\left| V\right\rangle )\}$ and for Bob, $%
\{\left| H\right\rangle ,\left| V\right\rangle ,\left| D\right\rangle =\frac{%
1}{\sqrt{2}}(\left| H\right\rangle +\left| V\right\rangle ),\left|
L\right\rangle =\frac{1}{\sqrt{2}}(\left| H\right\rangle -i\left|
V\right\rangle )\}$, respectively, and these 16 kinds of the joint
projective measurements were made for the linear tomography\cite%
{WhitePRL,Usami}. In order to reduce background noise and select the
frequency-degenerate bi-photons at 532 $%
%TCIMACRO{\unit{nm}}%
%BeginExpansion
\mathop{\rm nm}%
%EndExpansion
$, we employed interference filters centered at 532 $%
%TCIMACRO{\unit{nm}}%
%BeginExpansion
\mathop{\rm nm}%
%EndExpansion
$ (bandwidth (FWMH) of 8.0 $%
%TCIMACRO{\unit{nm}}%
%BeginExpansion
\mathop{\rm nm}%
%EndExpansion
$, peak transmissivity of 52\%) in front of the detectors. 

\begin{figure}
\epsfxsize=7.0cm
\centerline{\epsfbox{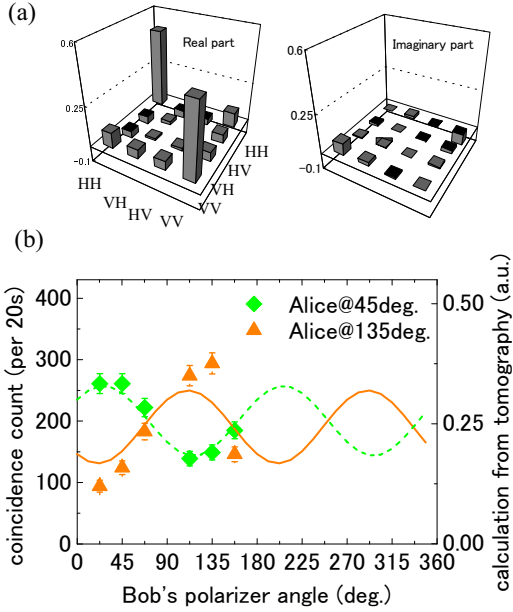}}
\caption{(a) The density matrix numerically reconstructed by maximum
likelihood method (MLM) (the eigenvalues are \{0.606, 0.388, 0.006, 0\}%
\protect\cite{WhitePRL,Usami,James}. Note that without MLM, we obtained the 
{\it illegitimate} density matrix (the eigenvalues are \{0.641, 0.361,
0.080, -0.082\}). (b) The results of polarization-correlation measurements
for checking Bell-CHSH inequality. The points represent the measured
coincidence counts and the lines represent the calculated coincidence rate
derived from the reconstructed density matrix. Only the results of
polarization-correlation measurements where Alice's polarizer angle \ is set
to be 45$%
%TCIMACRO{\unit{\U{b0}}}%
%BeginExpansion
\mathop{\rm %
{{}^\circ}}%
%EndExpansion
$ and 135$%
%TCIMACRO{\unit{\U{b0}}}%
%BeginExpansion
\mathop{\rm %
{{}^\circ}}%
%EndExpansion
$ are shown, though for checking CHSH inequality, other results ( 0$%
%TCIMACRO{\unit{\U{b0}}}%
%BeginExpansion
\mathop{\rm %
{{}^\circ}}%
%EndExpansion
$ and 90$%
%TCIMACRO{\unit{\U{b0}}}%
%BeginExpansion
\mathop{\rm %
{{}^\circ}}%
%EndExpansion
$ ) were used. }
\label{fig3}
\end{figure}

Figure\ref{fig3}%
(a) shows the reconstructed density matrix under this spectral filtering
condition. Note, the matrix which is linearly reconstructed by using only
tomography is not always legitimate, i.e., positive definite. We thus need
numerical optimization, called maximum likelihood method for obtaining a
legitimate and reliable density matrix\cite{James}. From this numerically
reconstructed density matrix, we calculated concurrence (ref.\cite{Wootters}%
) and von Neumann entropy ($\sigma =-Tr\left[ \rho _{AB}\;\ln \rho _{AB}%
\right] $ $(0\leq \sigma \leq \ln 4\simeq 1.39)$, where $\rho _{AB}$ stands
for the density matrix of principal system). We obtained concurrence of
about 0.21 and von Neumann entropy of about 0.70. This disappointingly poor
entanglement and purity might be due to the aforementioned which-crystal
information. For checking whether this state violate Bell-CHSH inequality (S%
%TCIMACRO{\TEXTsymbol{<}}%
%BeginExpansion
\mbox{$<$}%
%EndExpansion
2)\cite{CHSH} or not, we independently performed polarization-correlation
measurements\cite{BellEX} (Fig.\ref{fig3}(b)). We obtained the value S in
Ref.\cite{BellEX} of about 1.6. Thus, the bi-photon state under above
spectral filtering condition does not violate Bell-CHSH inequality, but it
nevertheless possesses entanglement because of non-zero concurrence. This
fact implies that Bell inequality is not a good measure for evaluating
entanglement\cite{HiddenNonlocal,Usami}. Figure\ref{fig3}(b) shows the
measured coincidence counts (points) and the calculated coincidence rate
from the reconstructed density matrix (line). Their good agreement confirms
the reliability of our reconstructed density matrix.

\begin{figure}
\epsfxsize=7.0cm
\centerline{\epsfbox{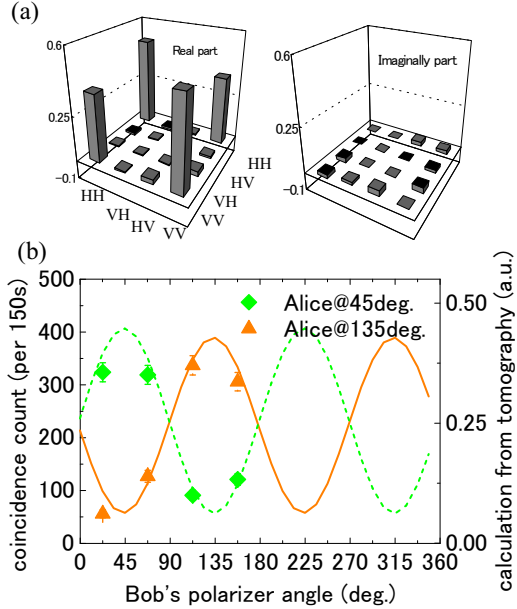}}
\caption{(a) The density matrix reconstructed by maximum likelihood method
(the eigenvalues are {0.869, 0.131, 0, 0}). (b) The results of
polarization-correlation measurements for checking Bell-CHSH inequality.}
\label{fig4}
\end{figure}

Next, in order to erase which-crystal information, we employed narrow
spectral filters in front of the detectors (bandwidth (FWMH): 1.2 $%
%TCIMACRO{\unit{nm}}%
%BeginExpansion
\mathop{\rm nm}%
%EndExpansion
$; peak transmissivity: 33\%). Figure\ref{fig4}(a) shows the density matrix
numerically reconstructed from the experiment. As expected, the off-diagonal
parts of the density matrix were retrieved, and we obtained its concurrence
of about 0.74 and von Neumann entropy of about 0.39. Furthermore, from the
polarization-correlation measurement (Fig.\ref{fig4}(b)), Bell-CHSH's value,
S in Ref.\cite{BellEX}, is about 2.4 which indicates the violation of
Bell-CHSH inequality. Therefore the results show that we have succeeded in
restoration of entanglement from the so-called hidden non-local state\cite%
{HiddenNonlocal,KwiatNature} ($C\sim 0.21$) to the explicitly non-local
state ($C\sim 0.74$) by a disentanglement eraser.

For theoretically estimating how much entanglement should be restored by our
disentanglement erasing model, let us go back to Eq.\ref{3 system}. To
obtain the reduced density matrix of the principal system alone, we perform
a partial trace over the tag system, then we get $\rho _{AB}=Tr_{T}\left[
\left| \Omega _{ABT}(t)\right\rangle \left\langle \Omega _{ABT}(t)\right| %
\right] =\int \langle \omega ^{\prime }\left| \Omega _{ABT}(t)\right\rangle
\left\langle \Omega _{ABT}(t)\right| \omega ^{\prime }\rangle d\omega
^{\prime }$. It can be rewritten as 
\begin{eqnarray}
\rho _{AB} &=&\frac{1}{2}\{\left| H\right\rangle _{A}\left| H\right\rangle
_{B}\left\langle H\right| _{A}\left\langle H\right| _{B}+\left|
V\right\rangle _{A}\left| V\right\rangle _{B}\left\langle V\right|
_{A}\left\langle V\right| _{B}  \nonumber \\
&&+C\left| H\right\rangle _{A}\left| H\right\rangle _{B}\left\langle
V\right| _{A}\left\langle V\right| _{B} \nonumber \\
&&+C^{\ast }\left| V\right\rangle
_{A}\left| V\right\rangle _{B}\left\langle H\right| _{A}\left\langle
H\right| _{B}\},  \label{reduced density matrix}
\end{eqnarray}%
where $C=\int \langle \omega ^{\prime }\left| \psi _{1}(t)\right\rangle
\left\langle \psi _{2}(t)\right| \omega ^{\prime }\rangle d\omega ^{\prime }=%
\frac{1}{(2\pi )^{2}}\int \left| g(\omega ^{\prime })\right| ^{2}e^{i\omega
^{\prime }\tau }d\omega ^{\prime }$, and according to the Wiener-Khintchin
theorem, we get%
\begin{equation}
C=\frac{1}{2\pi }\int \psi _{1}^{\ast }(t+\tau )\psi _{1}(t)dt.
\label{overlap}
\end{equation}%
Here we define the complex function in Eq.\ref{overlap} as $\psi _{1}(t)=%
\frac{1}{2\pi }\int g(\omega _{i})e^{-i\omega _{i}t}d\omega
_{i}=A(t)e^{i\theta (t)}$, where $A(t)$ and $\theta (t)$ respectively
represent the envelope of the amplitude and the phase (including the chirp
term) of the bi-photon state. Since the concurrence of the reduced density
matrix of principal system expressed as Eq.\ref{reduced density matrix} is
exactly the C of Eq.\ref{overlap}, all we have to do is to estimate the C.
We obtain C=0.63 for the spectral filters of 8.0 $%
%TCIMACRO{\unit{nm}}%
%BeginExpansion
\mathop{\rm nm}%
%EndExpansion
$-bandwidth and C=0.99 for the spectral filter of 1.2 $%
%TCIMACRO{\unit{nm}}%
%BeginExpansion
\mathop{\rm nm}%
%EndExpansion
$-bandwidth with the following three assumptions: (1) the pulse shape of
wave packet $\left| \psi _{1}(t)\right\rangle $ is Gaussian centered at $t=0$
, whose pulse duration determined by the bandwidth of the spectral filters;
(2) this Gaussian pulse is transform-limited; (3) the time difference $\tau
\simeq 100$ $%
%TCIMACRO{\unit{fs}}%
%BeginExpansion
\mathop{\rm fs}%
%EndExpansion
$. Thus this calculated results are consistent with our experiment. The
difference between the value derived by our theoretical model and that
obtained by experiment might be due to other distinguishabilities of the
entangled states in the experiment and the assumptions in the theoretical
estimation.

It is worthy to mention that the disentanglement eraser decreased the von
Neumann entropy of the principal systems. On the other hand, in the recent
experiment of the entanglement distillation using Procrustean method\cite%
{KwiatNature}, von Neumann entropy of the systems increased. 

In summary, we have shown that entanglement of the bi-photon polarization
state generated by pulse-pumped SPDC can be restored by the narrow-bandwidth
spectral filtering. We found that spectral filtering effectively erased the
possibilities of distinguishing the two bi-photon wave packets; thus,
entanglement of the principal system (polarization state) can be restored.
This erasing procedure will open up the alternative possibility of
entanglement manipulation and quantum information processing.

We gratefully acknowledge useful discussion with S. Kono, B. -S. Shi, and A.
Tomita.


\begin{references}
\bibitem{QCQI} M. A. Nielsen and I. L. Chuang, {\it Quantum Computation and
Quantum Information}, (Cambridge University Press, 2000)

\bibitem{CHSH} J. F. Clauser, {\it et al}., Phys. Rev. Lett. {\bf 23}, 880
(1969)

\bibitem{BellEX} A. Aspect, P. Grangier, and G. Roger, Phys. Rev. Lett. {\bf %
49}, 91 (1982); G. Weihs, {\it et al}., Phys. Rev. Lett. {\bf 81}, 5039
(1998)

\bibitem{Crypto} A. K. Ekert, Phys. Rev. Lett. {\bf 67}, 661 (1991); T.
Jennewein, {\it et al}., Phys. Rev. Lett. {\bf 84}, 4729 (2000); D. S. Naik, 
{\it et al}., Phys. Rev. Lett. {\bf 84}, 4733 (2000); W. Tittel, {\it et al}%
., Phys. Rev. Lett. {\bf 84}, 4737 (2000)

\bibitem{QTtheo} C. H. Bennett, {\it et al}., Phys. Rev. Lett. {\bf 70},
1895 (1993)

\bibitem{QTInsEX} D. Bouwmeester, {\it et al}., Nature (London) {\bf 390},
575 (1997)

\bibitem{QTEX} D. Boschi, {\it et al}., Phys. Rev. Lett. {\bf 80}, 1121
(1998); A. Furusawa, {\it et al}., Science {\bf 282}, 706 (1998); Y. -H.
Kim, S. P. Kulik, and Y. Shih, Phys. Rev. Lett. {\bf 86}, 1370 (2001)

\bibitem{QDenCode} C. H. Bennett and S. J. Wiesner, Phys. Rev. Lett. {\bf 69}%
, 2881 (1992); K. Mattle, {\it et al}., Phys. Rev. Lett. {\bf 76}, 4656
(1996)

\bibitem{RevESPDC} Y. -H. Kim, {\it et al}., Phys. Rev. A {\bf 63}, 062301
(2001) and reference therein

\bibitem{KwiatPRA} P. G. Kwiat, {\it et al}., Phys. Rev. A {\bf 60}, R773
(1999)

\bibitem{WhitePRL} A. G. White, {\it et al}., Phys. Rev. Lett. {\bf 83},
3103 (1999)

\bibitem{KwiatScience} P. G. Kwiat, {\it et al}., Science {\bf 290}, 498
(2000)

\bibitem{KwiatNature} P. G. Kwiat {\it et al}., Nature (London) {\bf 409},
1041 (2001)

\bibitem{Zukowski} M. Zukowski, {\it et al}., Phys. Rev. Lett. {\bf 71},
4287 (1993); M. Zukowski, A. Zeilinger, and H. Weinfurter, Annals of the
N.Y. Acad. of Sciences {\bf 755}, 91 (1995)

\bibitem{EnManipuEX} J. -W. Pan, {\it et al}., Phys. Rev. Lett. {\bf 80},
3891 (1998); J. -W. Pan, {\it et al}., quant-ph/0104047; D. Bouwmeester, 
{\it et al}., Phys. Rev. Lett. {\bf 82}, 1345 (1999); J. -W. Pan {\it et al}%
., Nature (London) {\bf 403}, 515 (2000)

\bibitem{Garisto-Hardy} R. Garisto and L. Hardy, Phys. Rev. A {\bf 60}, 827
(1999)

\bibitem{Wootters} W. K. Wootters, Phys. Rev. Lett. {\bf 80}, 2245 (1998)

\bibitem{Usami} K. Usami, Ms. of Engineering thesis, Tokyo Institute of
Technology, Japan (2001)

\bibitem{James} D. F. James, {\it et al}., quant-ph/0103121

\bibitem{KimPRArapid} Y. -H. Kim, S. P. Kulik, and Y. Shih, Phys. Rev. A 
{\bf 63}, 060301(R) (2001)

\bibitem{Scully} M. O. Scully, B. -G. Englert, and H. Walther, Nature
(London) {\bf 351}, 111 (1991)

\bibitem{SpectFilter} G. Di Giuseppe, {\it et al}., Phys. Rev. A {\bf 56},
R21 (1997); W. P. Grice, {\it et al}., Phys. Rev. A {\bf 57}, R2289 (1998)

\bibitem{NMR-eraser} G. Teklemariam, {\it et al}., Phys. Rev. Lett. {\bf 86}%
, 5845 (2001)

\bibitem{Braunstein} S. L. Braunstein, {\it et al}., Phys. Rev. Lett. {\bf 83%
}, 1054 (1999)

\bibitem{HiddenNonlocal} R. F. Werner, Phys. Rev. A {\bf 40}, 4277 (1989);
S. Popescu, Phys. Rev. Lett. {\bf 72}, 797 (1994)
\end{references}
\end{document}